# Simultaneous recording of two- and four-probe resistive transitions in doped laser-processed Sr-Ru-O

A. M. Gulian[1*], V. R. Nikoghosyan[2], D. Van Vechten[3], and K. S. Wood[4]

Abstract:

To confirm previously reported evidence of high-temperature superconductivity in laser processed Sr-Ru-O, we performed simultaneous two-probe and four-probe resistive measurements using bar-geometry samples. A superconducting-type transition with an onset at about 250K was recorded in one of the samples, consistent with our previously reported measurements in the X-bridge geometry. Some compositional details of the samples are also provided which were not known at the time of previous web-publication.

[1)] Physics Art Frontiers, Ashton, Maryland, 20861-9747, USA
[2)] Physics Research Institute, Ashtarak, 378410, Armenia
[3)] Office of Naval Research, Arlington VA, 22203-1995, USA
[4)] Naval Research Laboratory, Washington DC, 20375-0001, USA

*Corresponding author:  gulian@hotmail.com

## 1. Introduction
In the previous communication [1] multiple classes of experimental findings were reported which support the hypothesis that superconductivity occurs above 200K in the laser-processed doped Sr-Ru-O system. These included four-probe resistive transitions to an apparent R=0 state. While such observations are a necessary condition for presuming superconductivity, four-probe resistance measurement have a drawback: in samples with dimensionality higher than one, current can, in principle, flow along a trajectory that does not produce a potential between the voltage readout leads. The simplest example is a crack that opens while cooling (illustrated in Fig. 1), creating misleading results.

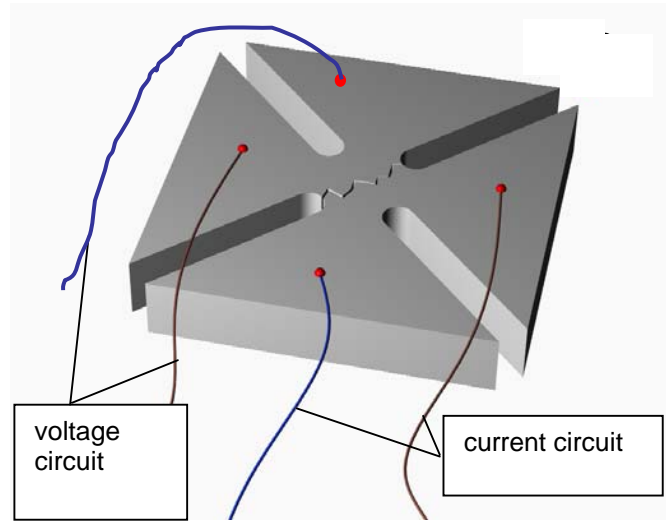

Figure 1. For samples with the X-geometry reported in [1], cracking could create artifacts that look like "superconducting transitions".

For this reason, measurements using a four-probe technique are not free of questions. Two-probe measurements are considered[1] more rigorous. While having the disadvantage of the added contact resistance, they have a benefit of measuring the voltage difference along the exact current trajectory. Simultaneous two and four probe measurements thus are highly desirable to complement each other.

## 2. Sample preparation
A new sample for two-probe measurements was prepared in a bar geometry (Fig. 2). It was prepared using a thin non-polished (cleaved) plate of bulk $Sr_2RuO_4$ crystals[2] with thickness about 100μm. The same metallic layers were deposited and laser-processed as described in Ref. [1]. Careful EDX analysis of our initial samples revealed that there is a thin Ni-Fe-Cr (nichrome) layer beneath of the Ag-layer[3]. The makers of the laser-processed samples had deposited this layer in order to increase the adhesion of the silver films to the crystals. The nominal composition of the nichrome deposition target was: Ni~10%, Fe~68%, Cr~17%, and 3%-other impurities[4], and nominal thickness of deposited layer was 40-70nm.

---
[1] We are grateful to Dr. T. Geballe and the participants of his seminar at Stanford for suggesting the idea of two-probe measurements.
[2] We are grateful to Dr. Y. Maeno (Kyoto U.) for providing high-quality single-crystals.
[3] We are indebted to Dr. S.-F. Cheng (NRL) for this finding.

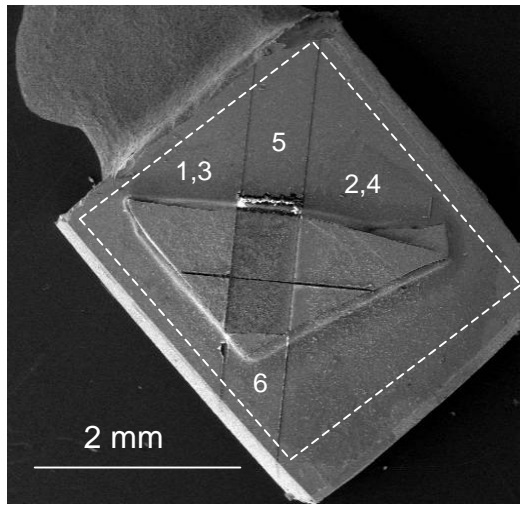
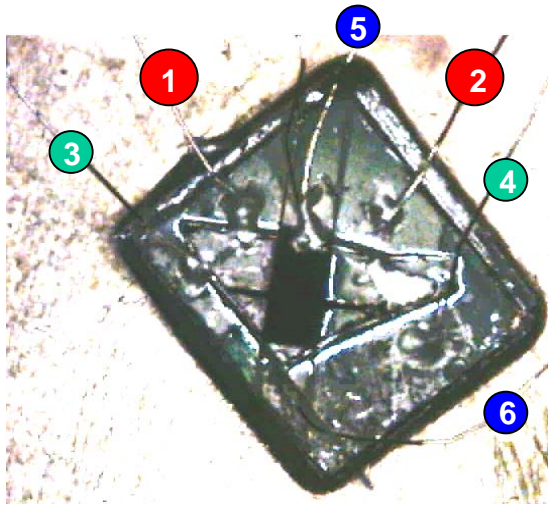
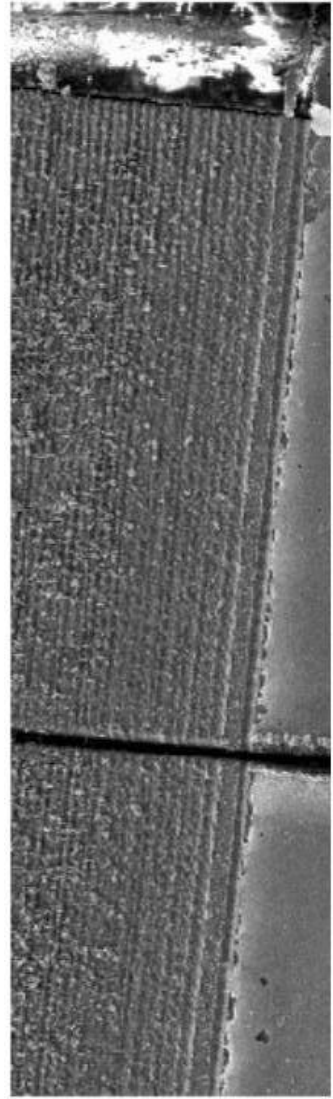

**Fig. 2.** The sample L13C128G1 in a bar geometry: photomicrographs and lead arrangement for simultaneous four-probe and two-probe measurements: current leads (1 and 2) and voltage leads (3 and 4) are connected correspondingly via high conductance silver film yielding two-probe measurements, leads (5 and 6) are connected through the sample only and provide (jointly with 1 and 2) four-probe measurement circuit.

It is explained in Ref. [1] that the original intention was not to produce or investigate high temperature superconductivity, but rather to pursue a specific applied program [2] utilizing low-$T_c$ triplet-state superconductivity, known in Sr-Ru-O since the 1990s [3]. For that original aim, deposition of this nichrome layer to assure adhesion was appropriate. However, for the purpose now at hand, that of assessing high-$T_c$ superconductivity in the resulting samples, it is a complicating factor. Great care had been taken to avoid contamination with magnetic impurities, and the presence of nichrome initially went unrecognized. Now that it has become recognized one must ask, as was already asked with regard to the overlying Ag deposition, whether the nichrome is somehow important to the outcome. Its presence certainly impacted both the magnetic measurements and interpretations. Both questions must be deferred until the high-$T_c$ superconductivity is proven beyond reasonable doubt.

In the new sample shown in Fig. 2, the laser-processed part has furrows in the vertical (longer) direction. Horizontally, there is a groove, which did not go all the way through the sample thickness. The groove was placed there to recreate all the conditions present in the original samples. In the X-geometry of the initially-reported samples [1], there were four half-diagonal grooves towards the center (see Fig. 1). Given the X-geometry, they could have contributed to, or have been wholly responsible for, the resistive transition allowing the super-current to flow just in the walls of the grooves and create zero voltage output. Moreover, the laser plume obviously had maximal influence at the walls inside the groove and modified more strongly the properties of material there. Since the discovery of transitions was accidental in the initial samples, departing minimally from the conditions used to produce those samples maximizes the chance of reproducing the original properties. Additional laser cuts formed four silver contact pads connected via the sample (Fig. 2, *top left*): the left and the right triangle pads (marked as 1,3 and 2,4 correspondingly), and the top (5) and bottom (6) rectangular pads in the middle. The area outside the dashed line was mechanically scratched off to secure connection between all the pads solely through the sample. Contact leads were arranged using silver paint as shown in the Fig. 2, *left bottom*.

**3. Resistive measurements**
For measurements we used 1 mA DC-current, reversing its polarity every 12 seconds, so that possible thermoelectric effects could be excluded by digital subtraction of negative from positive branches of voltage output. Voltage was measured in a longitudinal geometry between pads 3 and 4 in the Fig. 2, *left* panels (which constituted two-probe measurement with the current applied via pads 1 and 2) and in a transverse geometry between pads 5 and 6 (which constituted a four probe measurement with the current applied via pads 1 and 2). It is crucial that these measurements were taken simultaneously, *i.e.*, using multiple voltmeter circuits on the same thermal cycling run. The results are shown in Fig. 3.

---

[4] Weight percents everywhere.

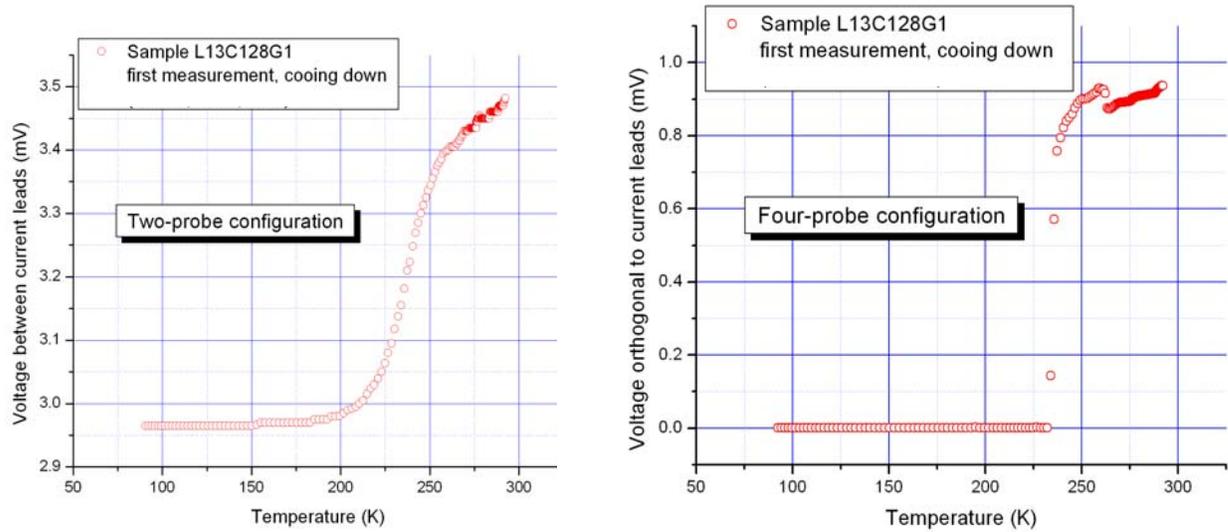

**Fig. 3.** Transitions in the two-probe (*left*) and four-probe (*right*) configurations at the first cooling. The transitions are recorded *simultaneously*.

## 4. Discussion

Let us consider the records from the sample being reported here in a greater detail and make sure that the transition in the two-probe circuit is not related to the contacts. Notice first that the *onset of the resistive transitions* in both circuits took place simultaneously (and therefore at the same T) so that the transitions are most likely caused by the same mechanism. Next, as follows from the Fig. 3 (*left*), above the transition the voltage value between the current leads in the two-probe circuit is about 3.5mV. This value is caused by the sample, by the contacts between the sample and the leads, and by the leads themselves. It drops at the transition by 0.5mV. In contrast, the four-probe voltage (Fig. 3, *right*) is caused solely by the electric potential distribution within the sample. (The contacts in this case do not contribute because of high input impedance of nanovoltmeter.) Above the transition, the sample exhibits a potential difference of about 1 mV between pads 5 and 6 and drops virtually to zero after the transition. That means that the sample's portion of the two-probe voltage could easily be 0.5mV. That is, the lead and contact resistance are not the only significant contribution. Presuming the laser-processed part undergoes a resistive transition into the R=0 state (as the four-probe measurement suggests), the two-probe voltage should drop by a fraction of its initial value that cannot be predicted more accurately without the knowledge of the exact current distribution within the sample. Overall, there is a quantitative concordance between the amplitudes of four-probe and two-probe transitions.

Again, the fabrication techniques used were not designed to produce homogeneous samples but rather for sample thinning. Indeed inhomogeneity is obvious. The furrows stem from successive passes of the laser and are ~10 μm apart (10 μm is the step resolution of the moving table), while the laser was focused to a 3 μm spot. Pulse to pulse separation of laser shots down the furrows was about 2 μm. Thus the processed areas overlap substantially along, but not perpendicular to the furrows. This suggests more continuous superconducting paths are formed along the furrows. Note that the direction between pads 3 and 4 is perpendicular to the laser-created furrows (Fig. 2, *right* panel), while between 5 and 6 the direction is primarily along the furrows.

What happened at lower temperatures after the apparent zeroing of the four-probe voltage? As Fig. 4 indicates, the voltage amplitude at the four-probe measurement dropped by a factor of 300, but not to the noise level: some oscillations of the current are still observable. This is because the voltage pads have not been placed directly onto the processed surface, but rather at a small distance from it (especially the contact 6 (see Fig. 2)). Thus, as explained in Fig. 4 (*right*), after the transition the voltmeter should still sense a small drop of the potential along the normal area between the contacts and the processed region, as it does (Fig. 4, *center*).

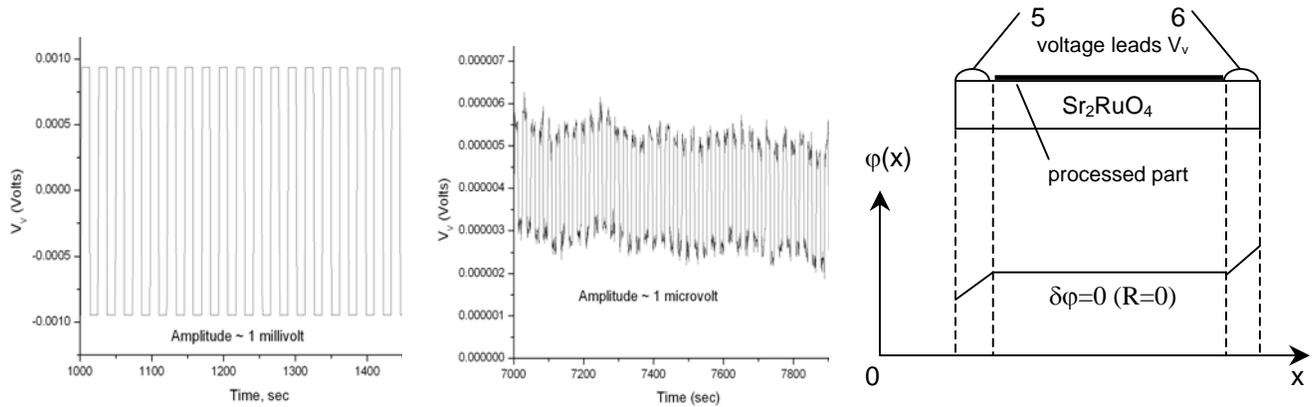

**Fig. 4.** Oscillating voltage is registered both above (*left*) and below (*center*) the transition. The *right* panel shows the pads arrangement and φ(x) distribution explaining the presence of non-zero voltage.

After the first cooling, at which the transitions were clearly seeable in both circuits, the subsequent warming reveal very different slope in two-probe circuit (Fig. 5, *left*).

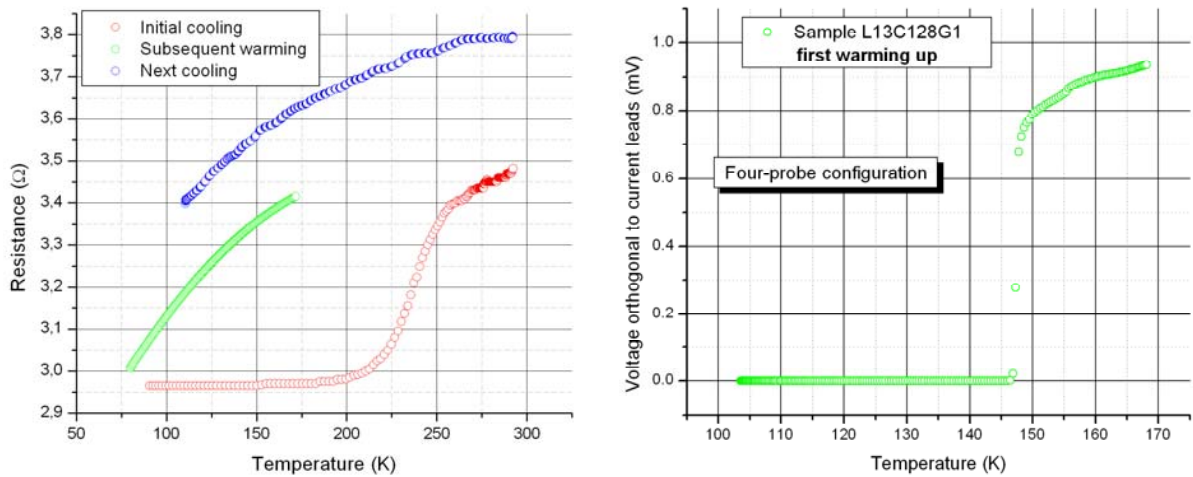

**Fig. 5.** Subsequent records with the sample L13C128G1 in the two-probe (*left* panel, green and blue curves) and four-probe (*right* panel) configurations.

That slope matches well with the slope of the second cooling curve if one moves downward the curve as a whole. Such a shift in the resistance may be associated with an additional, temperature independent resistance added, *e.g.*, by resistance changes in the silver-paint contacts). The hypothesis that these later curves are the result of the current passing almost entirely through the underlying host crystal rather than through the processed crust is tested in Fig. 6. The values of $\rho_{ab}$ and $\rho_c$ of $Sr_2RuO_4$ have very different temperature slopes [4] and differ at room temperature by a factor of 140. Thus they first were normalized by their values at the maximum temperature. Then the similarly normalized two-probe resistance data was fitted as a linear weight sum of the $\rho^n_{ab}$ and $\rho^n_c$, assuming temperature independent geometric factors/current trajectories. Fig. 6 shows the results, which clearly indicate the data is well fit by such a procedure and that the hypothesis is circumstantially confirmed. Thus, it appears the processed crust becomes electrically disconnected from the surrounding unprocessed regions (on which pads 1-4 reside) and current no longer flows through it. This is consistent with the hypothesis that the crust material down the furrows is much better interconnected than that perpendicular to the furrows.

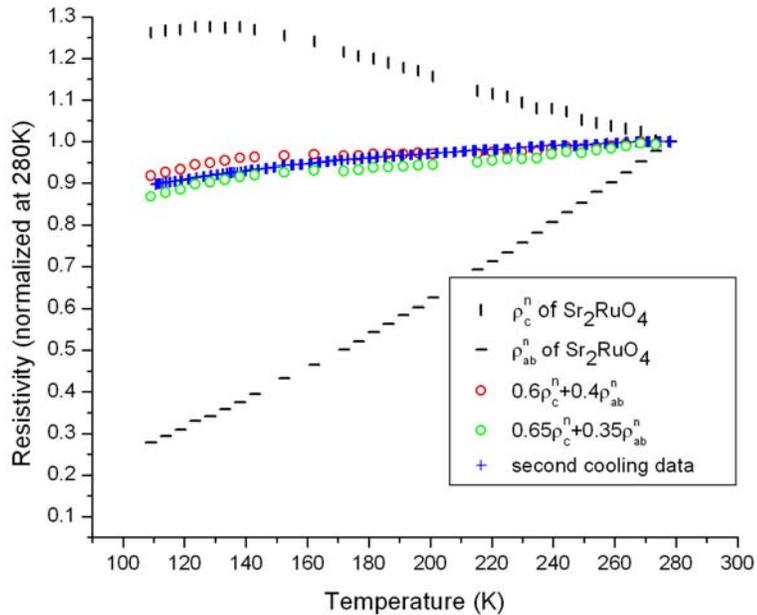

**Fig. 6.** Approximation for the second cooling curve in Fig. 5 by linear combination of $\rho_{ab}$ and $\rho_c$.

Thus it is not surprising that in the direction along the furrows (contact pads 5 and 6) the voltage drop is still shunted by possible superconductive current trajectories (Fig. 5, *right*) while no super-current flows in the perpendicular direction. It is likely the cracking that produced the two-probe disconnection also decreased the critical current along the furrows. That would explain the reduced apparent $T_c$ of 150K in the four-probe data (Fig. 5, *right*). (This fact initiated the stop of the recording at 170K.) The next run (second cooling) took place after a week-end. Upon resuming the records we were able to trace only one resistive channel (and chose the two-probe one, Fig. 5, *left, blue curve*). Subsequent runs showed larger additional series resistance so that the experiment was stopped.

Obviously, a single thermal cycle effectively destroyed this sample (with further damage on subsequent cycles), but its major success was to demonstrate simultaneous resistive transitions using the two-probe and four-probe techniques, which happened during the first cool down. The fact that the transition took place at the same temperature range as in the previously reported sample [1] is encouraging, since for that sample the transitions, though documented only by the four-probe technique, were repetitive.

## 5. Conclusions
In summary, resistive transitions have been registered during the first thermal cycle of a laser-processed $Sr_2RuO_4$ sample by simultaneous four-probe and two-probe techniques above 200K. Taken separately, the four-probe results can be explained by inter-sample switching of current trajectories (such as cracking). The two-probe transitions cannot be explained by cracks. Cracking should only enhance the resistance in this case. We also explained why the two-probe transition is not caused by the contact phenomena. The physical origins of the observed entirety of signals are of great interest in the context of high-temperature superconductivity.

**Acknowledgements**
This work was supported in part by the ONR Grant N0001403WX20850, by the ONR ROPO program, and by the NATO Grant SfP 974082.

We would like also to express our gratitude to many of our colleagues throughout the world for their help, support and interest in this work.